\documentclass[aip,amsmath,amssymb,preprint]{revtex4-2}

\usepackage{graphicx}
\usepackage{dcolumn}
\usepackage{bm}
\usepackage[utf8]{inputenc}
\usepackage[T1]{fontenc}
\usepackage{mathptmx}
\usepackage{etoolbox}
\usepackage{physics}
\usepackage{hyperref}
\usepackage{amssymb}

\makeatletter
\def\@email#1#2{%
 \endgroup
 \patchcmd{\titleblock@produce}
  {\frontmatter@RRAPformat}
  {\frontmatter@RRAPformat{\produce@RRAP{*#1\href{mailto:#2}{#2}}}\frontmatter@RRAPformat}
  {}{}
}%
\makeatother

\begin{document}

\title{Isolation of individual Er quantum emitters in anatase TiO$_2$ on Si photonics}


\author{Cheng Ji}
\altaffiliation{These authors contributed equally.}
\affiliation{Pritzker School of Molecular Engineering, University of Chicago, Chicago, IL 60637}
\affiliation{Materials Science Division, Argonne National Laboratory, Lemont, IL 60439}

\author{Robert M. Pettit}
\altaffiliation{These authors contributed equally.}
\affiliation{memQ Inc., Chicago, IL 60615}

\author{Shobhit Gupta}
\affiliation{memQ Inc., Chicago, IL 60615}

\author{Gregory D. Grant}
\affiliation{Pritzker School of Molecular Engineering, University of Chicago, Chicago, IL 60637}
\affiliation{Materials Science Division, Argonne National Laboratory, Lemont, IL 60439}

\author{Ignas Masiulionis}
\affiliation{Pritzker School of Molecular Engineering, University of Chicago, Chicago, IL 60637}
\affiliation{Materials Science Division, Argonne National Laboratory, Lemont, IL 60439}

\author{Ananthesh Sundaresh}
\affiliation{memQ Inc., Chicago, IL 60615}

\author{Skylar Deckoff--Jones}
\affiliation{memQ Inc., Chicago, IL 60615}

\author{Max Olberding}
\affiliation{memQ Inc., Chicago, IL 60615}

\author{Manish K. Singh}
\affiliation{memQ Inc., Chicago, IL 60615}

\author{F. Joseph Heremans}
\affiliation{Pritzker School of Molecular Engineering, University of Chicago, Chicago, IL 60637}
\affiliation{Materials Science Division, Argonne National Laboratory, Lemont, IL 60439}
\affiliation{Center for Molecular Engineering, Argonne National Laboratory, Lemont, IL 60439}


\author{Supratik Guha}
\affiliation{Pritzker School of Molecular Engineering, University of Chicago, Chicago, IL 60637}
\affiliation{Materials Science Division, Argonne National Laboratory, Lemont, IL 60439}
\affiliation{Center for Molecular Engineering, Argonne National Laboratory, Lemont, IL 60439}

\author{Alan M. Dibos}
\affiliation{Center for Molecular Engineering, Argonne National Laboratory, Lemont, IL 60439}
\affiliation{Q-NEXT, Argonne National Laboratory, Lemont, IL 60439}
\email{adibos@anl.gov}

\author{Sean E. Sullivan}
\affiliation{memQ Inc., Chicago, IL 60615}
\email{sean@memq.tech}

\begin{abstract}
Defects and dopant atoms in solid state materials are a promising platform for realizing single photon sources and quantum memories, which are the basic building blocks of quantum repeaters needed for long distance quantum networks. In particular, trivalent erbium (Er$^{3+}$) is of interest because it couples C-band telecom optical transitions with a spin--based memory platform. In order to produce quantum repeaters at the scale required for a quantum internet, it is imperative to integrate these necessary building blocks with mature and scalable semiconductor processes. In this work, we demonstrate the optical isolation of single Er$^{3+}$ ions in CMOS--compatible titanium dioxide (TiO$_2$) thin films monolithically integrated on a silicon--on--insulator (SOI) photonics platform. Our results demonstrate a first step toward the realization of a monolithically integrated and scalable quantum photonics package based on Er$^{3+}$ doped thin films.
\end{abstract}


\maketitle

The realization of long range quantum networks depends on the efficient distribution of entanglement across distant nodes. Quantum repeaters will be crucial for this task, allowing entanglement generation rates to overcome the exponential losses of optical fiber \cite{durQuantumRepeatersBased1999a, muralidharanUltrafastFaultTolerantQuantum2014, muralidharanOptimalArchitecturesLong2016}. Defects and dopant atoms in solid state materials are prime candidates for developing quantum repeaters \cite{aharonovichSolidstateSinglephotonEmitters2016, awschalomQuantumTechnologiesOptically2018}, and among them rare--earth ions are particularly attractive for their highly coherent optical and spin transitions \cite{thielRareearthdopedMaterialsApplications2011}. Notably, the trivalent erbium ion (Er$^{3+}$) offers a direct interface with telecom C--band optical photons, making it compatible with low--loss optical fiber networks. Er$^{3+}$ ions in solid state host crystals have been used to realize indistinguishable single photon emission \cite{ourariIndistinguishableTelecomBand2023a} as well as memories for quantum light  \cite{saglamyurekQuantumStorageEntangled2015a, weiQuantumStorage16502024}, which are the primary building blocks for quantum repeaters. 

An outstanding challenge for quantum repeaters is the development of a widely scalable platform that can be incorporated into mature semiconductor fabrication processes \cite{pettitPerspectivePathwayScalable2023}. For Er$^{3+}$, approaches so far have included flip--chip bonding of silicon waveguides and photonic crystal cavities onto doped bulk crystals \cite{ourariIndistinguishableTelecomBand2023a, uysalCoherentControlNuclear2023, dibosAtomicSourceSingle2018} and ion implantation directly into silicon \cite{kenyonErbiumSilicon2005, weissErbiumDopantsNanophotonic2021, gritschNarrowOpticalTransitions2022,gritschPurcellEnhancementSinglephoton2023a}. Another approach is to use the deposition of erbium doped host crystals onto silicon or SOI wafers either through molecular beam \cite{singhEpitaxialErdoped_22020, singhRareEarthBased2022, singhDevelopmentScalableQuantum2022, dibosPurcellEnhancementErbium2022, grantOpticalMicrostructuralCharacterization2024} or atomic layer deposition \cite{jiNanocavityMediatedPurcellEnhancement2024} techniques. Deposition allows for the exploration of host crystals compatible with growth on silicon that have other beneficial characteristics for hosting solid state defect centers such as wide bandgaps, low concentrations of nuclear spins and other impurities, as well as known site symmetries \cite{wolfowiczQuantumGuidelinesSolidstate2021}. For these reasons TiO$_2$ has been investigated as a promising host crystal for erbium, as titanium and oxygen have low natural abundances of isotopes with non--zero nuclear spin (12.85\% and 0.04\%, respectively), and Er$^{3+}$ ions predominantly take the substitutional Ti$^{4+}$ position in both of the rutile and anatase crystal phases which have a non--polar symmetry that prevents the formation of a permanent electric dipole moment \cite{phenicieNarrowOpticalLine2019, shinErdopedAnataseTiO2022}. In this work, we explore devices fabricated from anatase TiO$_2$ grown at lower temperatures than primarily rutile films. Previous work has shown that polycrystalline anatase phase TiO$_2$ grown on Si has smaller grain size and lower surface roughness\cite{singhRareEarthBased2022}, which is critical for nanophotonic cavity devices with low scattering loss needed to address individual Er$^{3+}$ ions.  

In this letter, we demonstrate that the deposition of Er:TiO$_2$ thin films on SOI wafers, subsequently patterned into nanophotonic waveguides and photonic crystal cavities, provides a platform capable of isolating single Er$^{3+}$ ions. At a temperature of 3.4 K, we tune the photonic crystal cavity within the inhomogeneous ensemble of ions and observe isolated peaks in photoluminescence excitation (PLE) scans with linewidths on the order of $\sim 100$~MHz, with a reduction of the optical emission lifetime by a factor >~400 and antibunching of the photon autocorrelation $g^{(2)}(0)<0.5$, revealing the emission of single photons. These results present Er:TiO$_2$ on SOI as a widely scalable platform that can be incorporated into mature semiconductor fabrication processes for the advancement of quantum technologies.  

\begin{figure}
    \includegraphics[width=0.98\textwidth]{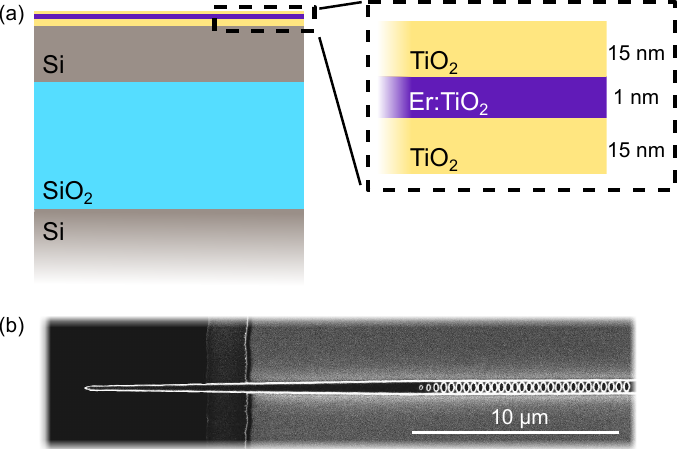}
    \caption{Overview of the platform to optically isolate single Er$^{3+}$ ions. (a) Outline of the device heterostructure. The heterostructure consists of a TiO$_{2}$ film grown on top of an SOI wafer with a 1 nm thick Er$^{3+}$ doped layer sandwiched between two 15 nm thick undoped layers. The doping level of the middle layer is 2 ppm. (b) SEM image of a representative fabricated device showing a 1D photonic crystal cavity embedded in a silicon optical waveguide.}
    \label{fig:one}
\end{figure}

Anatase TiO$_{2}$ thin films are grown on commercial SOI wafers (220~nm silicon layer with 2~$\mu$m buried oxide) and consist of three layers: a middle Er-doped layer sandwiched between undoped buffer and capping layers below and above, respectively\cite{singhRareEarthBased2022, singhDevelopmentScalableQuantum2022, dibosPurcellEnhancementErbium2022, solomonAnomalousPurcellDecay2024}. The doped layer has been shown to exhibit no significant erbium clustering \cite{singhRareEarthBased2022, jiNanocavityMediatedPurcellEnhancement2024}. In this study, we investigate a heterostructure in which the undoped layers are 15 nm thick and the doped layer is 1 nm thick.  The erbium concentration in the doped layer is estimated at 2 ppm based upon previous doping calibration confirmed with secondary ion mass spectrometry \cite{dibosPurcellEnhancementErbium2022, jiNanocavityMediatedPurcellEnhancement2024}. Figure \ref{fig:one}a provides an illustration of the TiO$_2$/SOI heterostructure.

In the crystal lattice, transitions within the unfilled $4f$ electronic orbital are weakly allowed and provide a direct C--band telecom optical interface near 1533 nm. Slight differences in the observed $Y_1$ $\rightarrow$ $Z_1$ crystal field transition in anatase TiO$_2$ have been reported for different growth conditions \cite{luoErDopedAnatase2011, shinErdopedAnataseTiO2022, jiNanocavityMediatedPurcellEnhancement2024}. Here, we observe the lowest energy PLE signature near 1532.8 nm which we probe in this work. Rare--earth ion optical transitions are long--lived, making weak photon emission signals from single ions hard to isolate without additional engineering. In order to enhance the emission rate, optical cavities can be used to modify the excited state lifetime of rare--earth ions through the Purcell effect \cite{martiniAnomalousSpontaneousEmission1987}. We therefore pattern waveguides and photonic crystal cavities into our SOI/TiO$_2$ heterostructure to enhance and collect the emission from ions in the thin film. We utilize a 1D photonic crystal cavity with elliptically shaped holes and a parabolic reduction of the lattice constant to confine the optical field to a mode volume of $\sim$~0.4 $(\lambda/n)^3$\cite{dibosPurcellEnhancementErbium2022}. Figure \ref{fig:one}b shows an SEM image of a representative device fabricated for this study. The photonic crystal cavity is embedded in a waveguide with a suspended inverse taper for efficient collection of the output light with a lensed optical fiber. The sample is placed in a closed--cycle cryostat at 3.4 K and the lensed optical fiber is aligned with 3--axis positioning stages. 

\begin{figure}
    \includegraphics[width=0.98\textwidth]{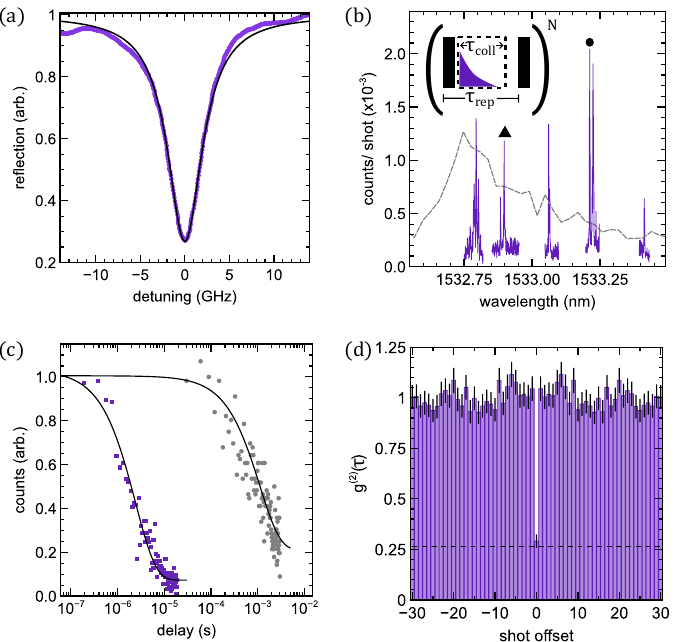}
    \caption{Purcell enhancement and antibunching of single photons. (a) The cavity reflection spectrum for the device used in this study. The black line is a Lorentzian fit to extract the cavity parameters. (b) PLE spectrum with resolved single ion emission lines that become visible when the cavity is tuned within the inhomogeneous line of the optical transition (gray dashed line, arbitrary scale). Inset: The optical pulse sequence used for PLE measurements. The duration of the signal collection window, $\tau_{\mathrm{coll}}$, is set independently from the repetition period of the pulse sequence, $\tau_{\mathrm{rep}}$. (c) The lifetime of the optical transition of a single ion resonant with the cavity (purple squares) is enhanced with respect to an ensemble in a waveguide with no cavity (gray circles). The single ion lifetime data is recorded for the ion marked with the triangle ($\blacktriangle$) in panel b. (d) A single photon autocorrelation on the ion marked with the circle ($\bullet$) in panel b, indicating single photon emission. The contribution of detector dark counts to the correlation is given by the dark shaded rectangles. The shot repetition period is 60~$\mu$s and the collection window is 20~$\mu$s. }
    \label{fig:two}
\end{figure}

In order to observe emission from single Er$^{3+}$ ions, we tune the resonance of a cavity across the inhomogeneous optical transition by freezing a thin layer of nitrogen gas on the surface of the sample via a closely placed capillary tube \cite{mosorScanningPhotonicCrystal2005}. Adsorption of the nitrogen leads to a redshift of the cavity mode which can be reversibly blueshifted through local heating of the cavity with resonant two--photon absorption by the Si. Figure \ref{fig:two}a shows the reflection spectrum for the cavity used in this work. The cavity exhibits a full--width at half--maximum linewidth of 4.7 GHz and a quality factor of 4.14$\times 10^4$. 

We perform PLE measurements, shown in Figure \ref{fig:two}b, by scanning a resonant laser within the cavity lineshape and observe discrete peaks in the emission recorded on a single photon detector (SNSPD), which we attribute to single Er$^{3+}$ ions. In the inset we show the pulse sequence used for recording PLE signal. After an optical pulse of duration 1.0~$\mu$s, signal from the ions is collected within a programmable time window of duration $\tau_{\mathrm{coll}}$, which can be set independently from the repetition period of the pulse sequence, $\tau_{\mathrm{rep}}$. The sequence is repeated $N$ times to build statistics. During the optical pulse we block the SNSPD with an acousto--optic modulator to prevent saturation of the detector signal.

The presence of the optical cavity on resonance with the ion induces a reduction in the optical lifetime due to the Purcell effect by a factor $P=\Gamma / \Gamma_0 - 1$, where $\Gamma=1/T_1$ is the observed decay rate and $\Gamma_0$ is the decay rate without the cavity. In Figure \ref{fig:two}c we show the observed lifetime of the ion near 1532.9~nm, marked with a triangle ($\blacktriangle$) in Figure \ref{fig:two}b. The observed lifetime is $T_1=$2.43(13)~$\mu$s, compared to 1.12(18)~ms measured for an ensemble of ions in a waveguide with no cavity on the same chip. This enhancement corresponds to a Purcell factor of $P=460$.

To confirm the emission of single photons from isolated Er$^{3+}$ ions we perform a second order photon autocorrelation measurement, shown in Figure \ref{fig:two}d for the ion near 1533.2~nm marked with a circle ($\bullet$) in Figure \ref{fig:two}b. We observe a zero delay value of $g^{(2)}(0)=0.29(3)$, which indicates single photon emission for $g^{(2)}(0)<0.5$. If we account for the background detection rate based on an independent measurement of the dark counts in our measurement (black dashed line), we observe an improved zero delay autocorrelation of $g^{(2)}(0)=0.04$. In this measurement the shot repetition period is 60~$\mu$s, providing a delay out to 1.8~ms for the 30--shot offset shown in Figure \ref{fig:two}d. The autocorrelation is symmetric about zero delay because the measurement is performed on a single detector.

\begin{figure}
    \includegraphics[width=0.98\textwidth]{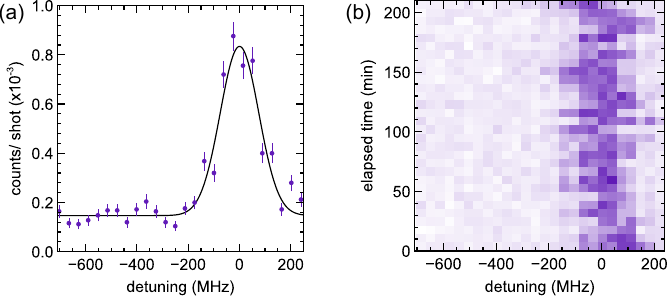}
    \caption{Single ion spectral diffusion. (a) A PLE scan across the single ion marked with a circle ($\bullet$) in Figure \ref{fig:two}b and fit with Gaussian lineshape. (b) Repeated PLE scans across the same ion over the course of $\sim3.5$~hrs. }
    \label{fig:three}
\end{figure}

In Figure \ref{fig:three} we characterize the linewidth and spectral diffusion of a single Er$^{3+}$ ion. We use the ion near 1533.2~nm again as a representative example. A single PLE scan reveals a Gaussian lineshape with a full--width at half--maximum of $\Delta\nu = 173.6$~MHz, indicating the presence of spectral diffusion with a timescale $\lesssim$ minutes required to perform the scan. This is in agreement with previous ensemble--based transient spectral hole burning measurements showing a few hundred MHz spectral diffusion linewidths on $\sim$10 ppm Er-doped anatase thin films\cite{singhDevelopmentScalableQuantum2022}. Over the course of $\sim$~3.5 hours, the ion displays additional slower spectral wandering, and the time--averaged linewidth increases to $\Delta\nu = 209.4$~MHz. The Gaussian lineshape indicates that the observed linewidth is not limited by pure dephasing. One route toward reducing this linewidth closer to the radiative limit, such that $\Delta\nu = 1/(2\pi \: T_1)$, is to further tailor the thickness of the buffering layers in the TiO$_2$ film to reduce the ion's proximity to interfaces. Additionally, we note that while anatase TiO$_2$ provides a non--polar symmetry at the titanium site, the site lacks inversion symmetry which makes the ion more susceptible to electric field noise \cite{sipahigilIndistinguishablePhotonsSeparated2014}. This indicates that rutile TiO$_2$ may be a better choice for hosting Er$^{3+}$, as the substitutional titanium site belongs to the inversion symmetric $D_{2h}$ point group. Preliminary transient spectral hole burning for Er in similar rutile phase TiO$_2$ devices has shown $\sim$50~MHz linewidths at much higher doping concentrations (>~100 ppm Er)\cite{solomonAnomalousPurcellDecay2024}. These results suggest that spectral diffusion is indeed reduced by switching to the rutile phase as host, and it is reasonable to assume single Er ion linewidths should be improved further by using the 2 ppm doping density in this current work. 

In this work we have demonstrated optical isolation of single Er$^{3+}$ ions in CMOS--compatible TiO$_2$ thin films monolithically integrated on an SOI photonics platform. Single ion linewidths have been observed on the order of $\sim$~100~MHz with Purcell enhancement factors of the optical decay rate >~400 and clear antibunching of the single photon emission. This work shows the promise of monolithically integrated Er:TiO$_2$ on SOI for the development of widely scalable approaches to incorporate quantum emitters with mature semiconductor fabrication processes. 

\section*{Acknowledgments}
\noindent
Authors R.M.P, S. Gupta., A.S., M.K.S., and S.E.S. acknowledge support from the U.S. Department of Energy Office of Science Advanced Scientific Computing Research program under Grant No. CRADA A22112 through the Chain Reaction Innovations program at Argonne National Laboratory. Authors C.J., G.D.G., I.M., S. Guha, and A.M.D. acknowledge the Q-NEXT Quantum Center, a U.S. Department of Energy, Office of Science, National Quantum Information Science Research Center, under Award Number DE-FOA-0002253 for support. Additional materials characterization support (F. J. H.) was provided by the U.S. Department of Energy, Office of Science; Basic Energy Sciences, Materials Sciences, and Engineering Division. All electron microscopy and device fabrication were performed at the Center for Nanoscale Materials, a U.S. Department of Energy Office of Science User Facility, supported by the U.S. DOE, Office of Basic Energy Sciences, under Contract No. DE-AC02-06CH11357. The authors would like to thank D. Czaplewski, C. S. Miller, R. Divan, and L. Stan for assistance with device fabrication.

\section*{Author Declarations}
\subsection*{Conflict of Interest}
\noindent
The authors have no conflicts of interest to disclose.


\bibliographystyle{quantum}
\bibliography{ErSingles}

\newpage

\end{document}